\documentclass[aps,prl,twocolumn,showpacs,preprintnumbers,amsmath,amssymb]{revtex4}
\usepackage{graphicx}
\usepackage{txfonts}

\begin{document}


\title{Thermal and electrical conductivity of Fermi pocket models of underdoped cuprates}
\author{M. F. Smith}
 \email{mfsmith@physics.uq.edu.au}
\affiliation{%
Department of Physics, University of Queensland,4072 Brisbane, Queensland, Australia}
\author{Ross H. McKenzie}
\affiliation{%
Department of Physics, University of Queensland,4072 Brisbane, Queensland, Australia}
\date{\today}

\begin{abstract}

Several  models of the electronic spectrum in the pseudogap state of underdoped cuprates have been proposed to explain ARPES and STM measurements, which reveal only truncated Fermi pockets instead of a full metallic Fermi surface.  We consider the transport properties expected of four physically distinct models, and calculate the thermal and electrical conductivity of the electronic quasiparticles.  By proposing transport currents that reflect the close correspondence between quasiparticles on the Fermi pockets in the pseudogap and those near nodes in the superconducting state, we show that measurable transport coefficients provide stringent tests of pseudogap models.

\end{abstract}


\maketitle

 Angle-resolved photoemission (ARPES) and scanning tunneling microscopy (STM) measurements on underdoped cuprates outside the superconducting dome have elucidated the electronic structure of the pseudogap state\cite{kond09,lee07,dama03,mars96,norm05,jlee09}. The large Fermi surface of overdoped systems is truncated to leave small arcs near nodal points.  Beyond the arcs, quasiparticles are gapped, with the gap growing larger towards the antinodal regions.  When the temperature is decreased below the superconducting transition, the arcs themselves become gapped by a $d$-wave superconducting order parameter\cite{jlee09,kohs08,push09}.

Transport measurements on underdoped systems complement the picture from ARPES and STM.  The thermal $\kappa(T)$ and electrical $\sigma(T)$ conductivity in the pseudogap state transitions smoothly into the superconducting state\cite{hawt03,doir06,suth05,prou06}, where the Bogoliubov quasiparticles are well-understood.   Wiedemann-Franz violations in the impurity-scattering regime, with the thermal conductivity exceeding the value expected from the Lorenz ratio $\kappa/(\sigma T)=L_0\equiv (k_B\pi)^2/3\mathrm{e}^2$, are seen.  Such violations are familiar in the $d$-wave superconducting state where fermionic quasiparticles in the superconductor carry heat and charge at different velocities\cite{durs00,abrik,aron80}.

Several models of the pseudogap state, reviewed by Norman et al.\cite{norm07}, introduce coherent parts of the single electron Green's function $G({\bf k},\omega)$ and can account for features of the ARPES data. (The origin and properties of these models, as well as their respective success in capturing the ARPES and tunneling spectra, are discussed in detail in Ref. \onlinecite{norm07} and references therein.)
Each model Green's function describes a metal with two coherent quasiparticle bands that intersect near nodal points with an electron distributed between the two bands according to fractional ${\bf k}$-dependent weights, just like in $d$-wave BCS theory.  Slave boson treatments of Hubbard-models, as well as simple mean field approaches,  have yielded Green's functions of this form\cite{ng05,lee06,lee09,hur09,yang06,yang09,metl10}.

The observable differences between the Green's functions for the candidate models are subtle, so the interpretation of additional experimental probes is crucial.  In this article, we study the quasiparticle transport properties of representative models.  Previous theoretical studies have obtained contradictory expressions for the quasiparticle electrical current--there is no obvious way to obtain a quasiparticle electrical current since its charge is not well-defined\cite{kiv90,lee98,shar03,kim02,vale07}.  We propose electrical and thermal currents that reflect the close correspondence between the observed quasiparticle transport in the pseudogap and $d$-wave superconducting states.  Using this approach, we find violation of the Wiedemann-Franz law and differences in the transport properties of models that would be difficult to distinguish using ARPES spectra alone.

Most of the single-electron Green's functions for the pseudogap models, reviewed in Ref. \onlinecite{norm07}, can be written:
\begin{equation}
\label{gf}
G({\bf k},\omega)=\sum_{\nu=+,-}\frac{W^2_{{\bf k},\nu}}{\omega-E_{{\bf k},\nu}}
\end{equation}
where
\begin{equation}
W^2_{{\bf k},\pm}=\frac{1}{2}\bigg{(}1\pm\frac{\xi_{\bf k}}{E_{\bf k}}\bigg{)},
E_{{\bf k},\pm}=\mu_{\bf k}\pm E_{\bf k},
\end{equation}
are the spectral weights of the $`+'$ and $`-'$ quasiparticle bands, $E_{\bf k}=\sqrt{\xi_{\bf k}^2+\Delta_{\bf k}^2}$ their respective energies, and $\mu_{\bf k}=(1/2)[\epsilon_{\bf k}+\epsilon^\prime_{\bf k}]$, $\xi_{\bf k}=(1/2)[\epsilon_{\bf k}-\epsilon^{\prime}_{\bf k}]$.  The pseudogap takes a simple $d$-wave form: $\Delta_{\bf k}=\Delta_0|\cos k_x - \cos k_y|$ while $\epsilon_{\bf k}$ and $\epsilon^\prime_{\bf k}$ are tight-binding band energy expressions where the distinct form chosen for $\epsilon^\prime_{\bf k}$ characterizes each model (see Table 1).
\begin{table}
\begin{tabular}{|c|c|c|c|c|} \hline
Model: & BCS & EDN & CDW & YRZ \\ \hline
$\epsilon^{\prime}_{\bf k}$: & $-\epsilon_{\bf k}$ & $-\epsilon_{\bf k} + \mu_0$ & $\epsilon_{\bf k+Q}$ & $-\epsilon_{0{\bf k}}$ \\ \hline
$\mu_{\bf k}$: & 0 & $\mu_0$ & $(1/2)[\epsilon_{\bf k}+\epsilon_{\bf k+Q}]$ & $(1/2)[\epsilon_{\bf k}+\epsilon_{0{\bf k}}]$ \\ \hline
\end{tabular}
\caption{Four models of the pseudogap state.  Each model has a single electron Green's function of the form Eq. \ref{gf} and accounts for some aspects of the ARPES spectra.  They differ in the form of $\epsilon^\prime_{\bf k}$.  The quantity $\epsilon_{\bf k}$ is a tight-binding band energy expression that includes several hopping terms while $\epsilon_{0{\bf k}}$ is restricted to nearest neighbour hopping.  The abbreviations are defined in the text.}
\end{table}

To calculate the appropriate transport coefficients we start by assuming that we can write down a free-fermion Hamiltonian with energies given by the poles of the Green's function:
\begin{equation}
\label{hdiag}
\hat{H}=\sum_{{\bf k},\sigma}H_{\bf k}=\sum_{{\bf k},\sigma}\bigg{(}E_{{\bf k},+}\gamma^{\dagger}_{{\bf k},\sigma,+}\gamma_{{\bf k},\sigma,+}+E_{{\bf k},-}\gamma^{\dagger}_{{\bf k},\sigma,-}\gamma_{{\bf k},\sigma,-}\bigg{)}
\end{equation}
where $\gamma_{{\bf k},\sigma,\pm}$ are quasiparticle operators.  The corresponding (free quasiparticle) matrix Green's function $\hat{G}_{qp}({\bf k},\omega)$ is diagonal, with respective elements: $(\omega-E_{{\bf k}+})^{-1}$ and $(\omega-E_{{\bf k}-})^{-1}$.

For the model Green's functions above, electrons are fractionally distributed between bands.  So quasiparticles will not carry charge at their band velocity and the scattering of electrons will couple the two quasiparticle bands (as is the case for superconductors \cite{abrik,aron80}).  To obtain simply heat and charge currents we transform to a basis of electrons, guided by the analogy with the $d$-wave superconductor.

 We thus assume a Bogoliubov transformation:
 \begin{equation}
  \hat{U} =\begin{bmatrix} W_{{\bf k},+} &  W_{{\bf k},-} \\ W_{{\bf k},-} & -W_{{\bf k},+} \end{bmatrix}
  \end{equation}
  which, when applied to the free-quasiparticle Green's function $\hat{G}_{qp}({\bf k},\omega)$ yields Eq. \ref{gf} as the $[11]$ matrix component.  Heat and charge currents are defined as for the quasiparticles in BCS, and transformed back into the quasiparticle basis.  The thermal velocity is taken to be
\begin{equation}
\label{vheat}
{\bf v}_{th}=\hat{U}\frac{d}{d\bf k}\bigg{(}\hat{U}\hat{H}_{\bf k}\hat{U}\bigg{)}\hat{U}=\begin{bmatrix} {\bf v}_{+} & {\bf \bar{v}} \\ {\bf \bar{v}} & {\bf v}_-\end{bmatrix}.
\end{equation}
where, in addition to the band velocities ${\bf v}_+$ and ${\bf v}_-$ there appears a quantity ${\bf\bar{v}}=-(\xi_{\bf k}/E_{\bf k})d\Delta_{\bf k}/d{\bf k}+(\Delta_{\bf k}/E_{\bf k})d\xi_{\bf k}/d{\bf k}$.  The latter mixes the contributions (to $\kappa(T)$) of the two quasiparticle bands in the presence of a phenomenological scattering rate $\gamma$.

For the electric charge velocity we use:
\begin{equation}
\label{vel}
{\bf v}_{el}=\hat{U}\begin{bmatrix} {\bf v}_f & 0 \\ 0 & {\bf v}^{\prime}_f \end{bmatrix}\hat{U},
\end{equation}
where ${\bf v}_f=d\epsilon_{\bf k}/d{\bf k}$ and ${\bf v}^{\prime}_f=d\epsilon^{\prime}_{\bf k}/d{\bf k}$.

We insert these expressions into the bubble diagrams for the thermal $\kappa/T$ and (quasiparticle) electrical conductivity $\sigma$:
\begin{equation}
\label{kap}
\frac{\kappa}{T}=\frac{1}{(2\pi)^2}\int \frac{d\omega\omega^2}{\pi T^2}\bigg{(}\frac{-df_0}{d\omega}\bigg{)}\int d^2{\bf k}\mathrm{Tr}\bigg{[}[G^{\prime\prime}_R({\bf k},\omega){\bf v}_{th}]^2\bigg{]}
\end{equation}
and
\begin{equation}
\label{sig}
\sigma=\frac{1}{(2\pi)^2}\int \frac{d\omega}{\pi}\bigg{(}\frac{-df_0}{d\omega}\bigg{)}\int d^2{\bf k}\mathrm{Tr}\bigg{[}[G^{\prime\prime}_R({\bf k},\omega){\bf v}_{el}]^2\bigg{]}
\end{equation}
where $G_R({\bf k},\omega)$ includes self-energy effects associated with scattering (here we assume scattering is momentum-independent so vertex corrections can be ignored).

Previous studies of the $d$-density-wave-, and BCS-derived models have disagreed\cite{kiv90,lee98,shar03,kim02,vale07} on the definition of the electrical current (Eq. \ref{vel}), particularly on whether to include off-diagonal elements associated with the velocity ${\bf v}_2\equiv d\Delta_{\bf k}/d{\bf k}$.  If these elements are included, then ${\bf v}_{th}$ and ${\bf v}_{el}$ are equal so Wiedemann-Franz violation does not occur\cite{shar03}. The observation of such violation and, more generally, the smooth evolution of quasiparticle transport across the superconducting transition\cite{doir06}, supports the assumption that, like in BCS, the electric current does not include the off-diagonal terms.  (Note that Eqs. \ref{vel} and \ref{vheat} generalize the results for $d$-wave superconductors\cite{notedurs} to cases where $\epsilon_{\bf k}\neq-\epsilon_{\bf k}^{\prime}$.)  We now contrast the transport properties of the models listed in Table 1.

 The CDW model uses $\epsilon^{\prime}_{\bf k}=\epsilon_{{\bf k}^\prime}$ where ${\bf k}^{\prime}={\bf k+Q}$ and ${\bf Q}=(\pi,\pi)$.  $\Delta_{\bf k}$ couples electrons at different ${\bf k}$-points in the full Brillouin zone, a quasiparticles is a combination of electrons at ${\bf k}$ and ${\bf k+Q}$.  The quasiparticle creation operator is: $\gamma_{{\bf k}\pm}=W_+ c_{{\bf k}}\pm W_-c_{{\bf k+Q}}$ and the appropriate electric velocities are then ${\bf v}_f=d\epsilon_{\bf k}/d{\bf k}$ and ${\bf v}_f^{\prime}=d\epsilon_{{\bf k}^{\prime}}/d{\bf k}^{\prime}$, which for nearest-neighbour hopping gives ${\bf v}_f^{\prime}=-{\bf v}_f$.  The two electronic components of the quasiparticle carry current in opposite directions.

 The BCS Green's function is obtained by taking $\epsilon^{\prime}_{\bf k}=-\epsilon_{\bf k}$.  A quasiparticle is built from the coupling of an electron at momentum  ${\bf k}$  and a hole of momentum ${\bf - k}$ and the quasiparticle are $\gamma_{{\bf k}\pm}=W_+ c_{{\bf k}}\pm W_-c^\dagger_{{\bf -k}}$.  Since both the velocity and charge of the quasiparticle's constituents differ in sign, they carry equal current so electric transport velocities satisfy  ${\bf v}_f={\bf v}^{\prime}_f$.  The sign change relative to the CDW case has crucial consequences for the Lorenz ratio, as seen below.  (Henceforth, we refer to the BCS-like relationship between the electron and hole velocities: ${\bf v}_f={\bf v}^{\prime}_f$ as electron-hole coupling, and to the CDW-like relationship ${\bf v}_f^{\prime}=-{\bf v}_f$ as electro-electron coupling.)

   For nearest neighbour hopping, the CDW and BCS models can be obtained by appropriate choice of parameters in another model discussed in Ref. \onlinecite{norm07}, the energy-displaced node model (EDN).  The band energies satisfy $\epsilon^{\prime}_{\bf k}=-\epsilon_{{\bf k}}$ and the parameter $\mu_{\bf k}=\mu_0$ is constant.  A finite $\mu_0$ provides for a finite zero-energy density of states (i.e. a Fermi pocket with a radius proportional to $\mu_0$) as illustrated in the inset of Fig 1.
\begin{figure}
\begin{center}
\includegraphics[width=4.0 in, height=4.0 in]{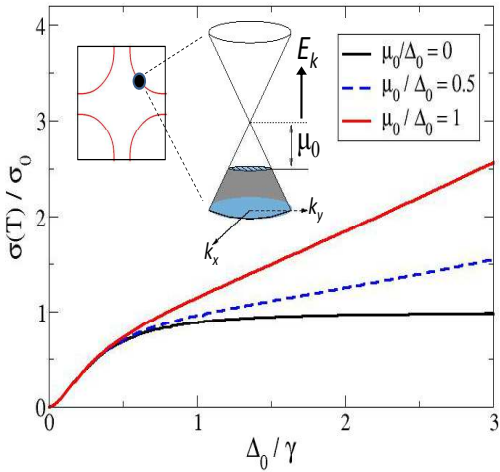}
\end{center}
\caption{\label{fig1} Fermi pockets, nodes and universal transport.  Several  model Green's functions of the pseudogap regime have quasiparticle dispersions resembling those in $d$-wave superconductors.  Near a node (indicated by the circle appearing along the zone diagonal on the normal state Fermi surface shown in the upper left) there are two intersecting quasiparticle bands, as illustrated.  A key feature of some of the models (including the EDN model, plotted here) is an energy shift, $\mu_0$, of the Fermi level from the node, which gives rise to a Fermi pocket.  The curves show the electrical conductivity $\sigma$ in units of the universal value $\sigma_0$ plotted versus scattering rate $\gamma/\Delta_0$ for different values of $\mu_0$.  When $\Delta_0/\gamma$ is large and quasiparticles are restricted to nodes, the value of $\mu_0$ determines whether quasiparticle transport behaves as in a normal metal (with conductivity proportional to $\gamma^{-1}$) or as in a $d$-wave superconductor (with universal conductivity $\sigma=\sigma_0$).  For small $\gamma/\Delta_0$, quasiparticles are excited everywhere on the Fermi surface and the parameter $\mu_0$ plays no role.}   \end{figure}
At low temperatures, $k_BT<<\max(\mu_0,\gamma)<<\Delta_0$ we find:
\begin{equation}
\label{kapmu}
\frac{\kappa}{T}=\frac{\kappa_0}{2T}\bigg{[}1+\bigg{(}\frac{\mu_0}{\gamma}+\frac{\gamma}{\mu_0}\bigg{)}\arctan\bigg{(}\frac{\mu_0}{\gamma}\bigg{)}\bigg{]}
\end{equation}
and, taking ${\bf v}_f^{\prime}=-{\bf v}_f$ as appropriate for the CDW model, we have
\begin{equation}
\label{sigmu}
\sigma=\sigma_0(\kappa/\kappa_0)
\end{equation}
where $\kappa_0/T=L_0[1+\alpha^2]^2$, where $\alpha\equiv v_2/v_f$ and $\sigma_0=(\mathrm{e}/\pi)^2\alpha^{-1}$. A similar result is obtained in Ref. \onlinecite{xia09}.
In the limit $\gamma>>\mu_0$ we recover the universal values for transport coefficients.  In the limit $\mu_0>>\gamma$ we find $(\kappa/T)\to(\kappa_0/T)[\pi\mu_0/4\gamma]$ and $\sigma\to\sigma_0(\mu_0\pi/4\gamma)$.  So the conductivities scale with the radius of the Fermi pocket $\mu_0$ and with the quasiparticle lifetime $1/\gamma$ as expected for a simple metal.  Note that $(L/L_0)=[1+\alpha^2]$ is independent of $\mu_0/\gamma$, so there is a constant enhancement of the heat conductivity relative to the Wiedemann-Franz prediction.  In Fig. 1 we show the conductivity in the EDN model with nearest neighbour hopping as a function of $\gamma$ for several values of $\mu_0$.  Evident is the crossover from behaviour at small $\mu_0$ to linear $\gamma^{-1}$-dependence.

While none of the models discussed in Ref. \onlinecite{norm07} considered a finite $\mu_0$ in the presence of an electron-hole coupling, it is simple and interesting to extend our analysis to such models.  Using ${\bf v}_f={\bf v}_f^{\prime}$ the thermal conductivity $\kappa$, Eq. \ref{kapmu}, is unchanged but the electrical conductivity becomes
$\sigma=\sigma_0\bigg{[}1+\frac{\mu_0}{\gamma}\arctan\bigg{(}\frac{\mu_0}{\gamma}\bigg{)}\bigg{]}$
which again gives $\sigma_0$ in the $\mu_0/\gamma<<1$ universal limit.  However, in the $\mu_0/\gamma>>1$ limit, gives a value twice as large as the corresponding limit of Eq. \ref{sigmu}.  (An analogous factor of two was found\cite{kim02} at finite temperature when $\mu_0=0$.)   So, for electron-hole couplings, $L/L_0$ is strongly $\gamma$-dependent and can be either smaller or larger than one.

 We now contrast the ARPES and transport predictions of the models discussed above.  In the CDW model, the predicted ARPES peaks trace out the elliptical Fermi pocket with axes of length $\mu_0/v_f$ and $\alpha\mu_0/v_f$.  (Because of the spectral weight, a portion of the back side of the pocket would be obscured, leaving only an arc.)    The corresponding transport prediction is that the conductivity should scale with $[\mu_0/\alpha] (1+\alpha^2)$ whereas the Lorenz ratio should be $L=L_0(1+\alpha^2)$.  If $\mu_0$ and $v_f/v_2$ changed differently with doping then one could impose a strong consistency check on the doping-dependence of the ARPES spectrum.

In the BCS model there are no pockets but there do appear arcs in the presence of a scattering rate $\gamma$ since the poles at energies $\pm E_{\bf k}$ are smeared together giving an arc of zero-energy spectral peaks along $\epsilon_{\bf k}=0$.  The length of the arc is proportional to $\gamma/v_2$ but the transport coefficents are universal, so changes in the scattering rate for a given size of the pseudogap would change the length of the arc without affecting the conductivity.

 As a concrete example of a possible doping evolution of the conductivities we consider the YRZ model\cite{hur09,yang06,yang09} obtained by taking $\epsilon^{\prime}_{\bf k}=-\epsilon_{0{\bf k}}=-2t[\cos k_x +\cos k_y]$, and $\epsilon_{\bf k}=-2t[\cos k_x + \cos k_y] -4t^{\prime}[\cos k_x \cos k_y]-2t^{\prime\prime}[\cos 2k_x + \cos 2k_y]+\mu_0$ where all the hopping coefficients and $\mu_0$ are dependent on doping in a precise manner prescribed by YRZ ($t^{\prime},t^{\prime\prime}$ and $\mu_0$ vanish at half-filling and increase linearly in magnitude with doping away from half-filling).  The resulting doping-dependence of the Fermi pockets is illustrated in the sketches on Figs. 2 and 3.  In addition, the weight of the quasiparticle is zero at half-filling and increases approximately linearly.

  The electric current velocities are given by ${\bf v}_f={\bf \nabla}_{\bf k}\epsilon_{\bf k}$ and ${\bf v}_f^\prime=\mp{\bf \nabla}_{\bf k}\epsilon_{0^{\bf k}}$ where the sign corresponds to the electron-electron (i.e. akin to CDW) and electron-hole (i.e. like BCS) coupling possibilities, respectively.

 If Fig. 2 we show the result for electron-hole coupling with the conductivity plotted in the main panel and the Lorenz ratio shown in the inset. The rapid increase of the conductivity with doping results from the growth of the Fermi pocket and the quasiparticle weight.  The Lorenz ratio is strongly doping dependent, especially when the scattering rate is low.  For small $\gamma$, $L/L_0$ attains large values at low doping (where the magnitude of the pseudogap is large), dips below one at intermediate doping and finally approaches one at large doping as the pseudogap vanishes.
\begin{figure}
\begin{center}
\includegraphics[width=4.0 in, height=3.6 in]{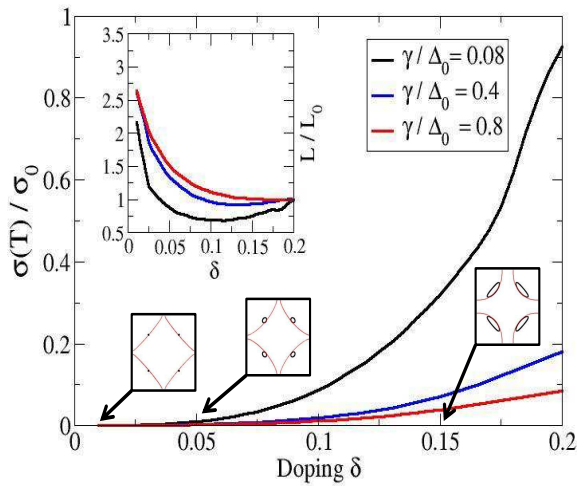}
\end{center}
\caption{\label{fig2} Doping-dependence of electrical and thermal conductivities. The electrical conductivity $\sigma/\sigma_0$ is plotted versus doping $\delta$ in the main panel, assuming the YRZ-evolution of electronic band parameters\cite{yang06,yang09}, and quasiparticle weight, for several values of the scattering rate $\gamma$ (in units of bare, doping-independent nearest-neighbour hopping parameter $t$).  The change of the Fermi pockets are illustrated by the small figures.  In the inset, the Lorenz ratio $L/L_0$, where $L_0$ is the Lorenz number, is plotted versus doping (showing deviations from the Wiedemann-Franz law $L=L_0$).  Values of $L/L_0$ significantly greater than one can occur at small $\delta$, owing to a large contribution to heat transport associated with momentum-variation of the pseudogap.  In clean systems (i.e. for small $\gamma$) with an electron-hole Bogoliubov transformation assumed (see text), Lorenz values of less than one are also possible. The qualitative doping dependence of the Lorenz ratio depends on model parameters, and can be used to distinguish candidate models (especially when the individual conductivities are not easy to interpret).} \end{figure}
In Figure 3, we contrast the case of electron-hole and electron-electron coupling, again using the doping dependence prescribed by the YRZ model.  The biggest difference between the two pairing scenarios is the strong $\gamma$-dependence seen in the electron-hole coupling but not the electron-electron coupling.

The existing thermal and electrical transport data on non-superconducting, underdoped cuprates supports the ARPES though may not yet rule out any models discussed\cite{hawt03,suth05,doir06,prou06}.  The quasiparticle contribution to conduction decreases with underdoping and large Lorenz values ($L/L_0>3$) have been seen in the field induced normal state in metallic, underdoped samples\cite{prou06}.  The Lorenz ratio is larger for dirtier samples (where the ratio of low and high temperature resistivies was used to estimate sample purity).  These results are in qualitative agreement with Fig. 3 and the dependence of the Lorenz ratio on scattering rate suggest that quasiparticles are formed from electron-hole coupling. However insulating behaviour has been reported that, in some studies, effects charge but not heat transport\cite{hawt03}, which would indicate a more radical departure between heat and charge currents than proposed above.
\begin{figure}
\begin{center}
\includegraphics[width=4.0 in, height=3.6 in]{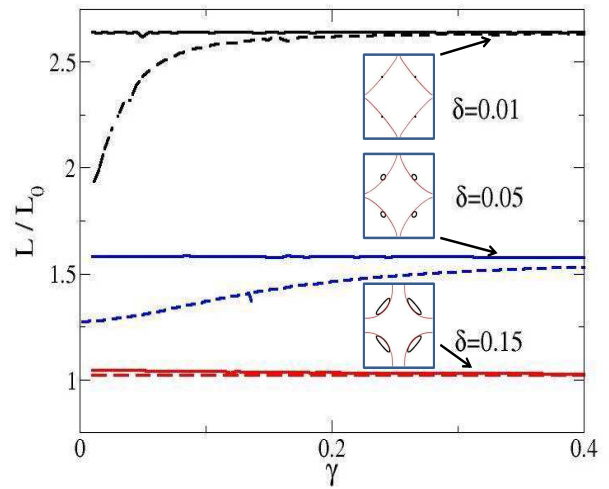}
\end{center}
\caption{\label{fig3} Scattering dependence of the Lorenz ratio in Fermi pocket models: contrasting electron-hole and electron-electron coupling.  The Lorenz ratio $L/L_0$ is plotted versus the scattering rate $\gamma$ for several values of doping $\delta$.  The doping dependence of all parameters is the same as in Fig. 2.  If the quasiparticles are formed from a combination of an electron and a hole (resulting in dashed curves in plot) then the Lorenz ratio is strongly dependent on scattering rate.  If they are formed from a pair of electrons at different momenta (resulting in solid curves) then the Lorenz ratio is independent of scattering rate.  The qualitative behaviour is significantly different between the electron-hole and electron-electron pairing scenarios (whereas both can give rise to the same single-electron spectral function as seen in ARPES data).  Experimental transport data on far-underdoped systems suggest a strong dependence of the Lorenz ratio on scattering rate in support of the electron-hole picture.} \end{figure}

In conclusion, transport coefficients of the model Green's functions, which describe coherent quasiparticles near nodes, offer tests of the models beyond those provided by the ARPES spectra that motivated them.  To combine the benefits of both probes, we have provided a plausible conjecture for the heat and charge currents that captures the observed connection between the pseudogap and the $d$-wave superconducting quasiparticles.  

This work was supported by Australian Research Council Discovery Projects DP1094395 and DP0710617.  We thank Ben Powell and X.-J. Xia for useful discussions.


\begin{thebibliography}{99}

    \bibitem{kond09}  T. Kondo, R. Khasanov, T. Takeuchi, J. Schmalian, and A. Kaminski,
Nature {\bf 457}, 296 (2009).
 \bibitem{lee07} W. S. Lee, I.M. Vishik, K. Tanaka, D. H. Lu, T. Sasagawa, N. Nagaosa,
T. P. Devereaux, Z. Hussain, and Z. X. Shen, Nature {\bf 450},
81 (2007).
\bibitem{dama03} A. Damascelli, Z. Hussain, Z.X. Shen, Rev. Mod. Phys. {\bf 75}, 473 (2003).

\bibitem{mars96} D. S. Marshall et. al., Phys. Rev. Lett. {\bf 76}, 4841 (1996);
H. Ding et. al.,
Nature {\bf 382}, 51 (1996); A. G. Loeser et al., Science {\bf 273}, 325 (1996);  A. Kanigel et al., Nat. Phys. {\bf 2}, 447 (2006).
\bibitem{norm05} M. R. Norman, D. Pines, and C. Kallin, Adv. Phys. {\bf 54}, 715
(2005).
\bibitem{jlee09} J. Lee et al., Science {\bf 325}, 1099 (2009).
\bibitem{kohs08} Y. Kohsaka, et al., Nature {\bf 454},
1072 (2008).
\bibitem{push09} A. Pushp et al., Science {\bf 324}, 1689 (2009).
\bibitem{hawt03} D. G. Hawthorn et al., Phys. Rev. Lett. {\bf 90} 197004 (2003).
\bibitem{doir06} N. Doiron-Leyraud, M. Sutherland, S.Y. Li, L. Taillefer, Ruixing Liang, D.A. Bonn, and W.N. Hardy, Phys. Rev. Lett. {\bf 97}, 207001 (2006).
\bibitem{suth05} M. Sutherland et al., Phys. Rev. Lett. {\bf 94}, 147004 (2005).
\bibitem{prou06} C. Proust, K. Behnia, R. Bel, D.Maude and S. I. Vedeneev, Phys. Rev. B {\bf 72}, 214511 (2005).
  \bibitem{durs00} A. C. Durst and P. A. Lee, Phys. Rev. B {\bf 62}, 1270 (2000).

 \bibitem{abrik} A. A. Abrikosov {\it Thoery of Normal Metals and Superconductors}, Elsevier (1992).
 \bibitem{aron80} A. G. Aronov, Y.M.Gal'perin, V. L. Gurevich, and V. I. Kozub, Adv. Phys. {\bf 30}, 539 (1981).


\bibitem{norm07} M. R. Norman, A. Kanigel, M. Randeria, U. Chatterjee, and J. C. Campuzano, Phys. Rev. B {\bf 76}, 175501 (2007).
\bibitem{ng05} T.-K. Ng, Phys. Rev. B {\bf 71}, 172509 (2005).

\bibitem{lee06} P. A. Lee, N. Nagaosa, and X.-G. Wen, Rev. Mod. Phys. {\bf 78}, 17
(2006).
\bibitem{lee09} T. Senthil and P. A. Lee, Phys. Rev. Lett. {\bf 103}, 076402 (2009).

\bibitem{xia09} X.-J. Xia and T.-K. Ng, J. Phy. Cond. Mat. {\bf 21}, 115703 (2009).

\bibitem{hur09} L. L. Hur and T. M. Rice Ann. of Phys. {\bf 324}, 1452 (2009)
\bibitem{yang06} K.-Y. Yang, T. M. Rice, and F.-C. Zhang, Phys. Rev. B {\bf 73},
174501 (2006)
\bibitem{yang09} K.-Y. Yang, H.-B. Yang, P. D. Johnson, T. M. Rice and F.-C. Zhang, Eur. Phys. Lett. {\bf 86}, 37002 (2009).
\bibitem{metl10} M. A.Metlitski and S. Sachdev, Phys. Rev. B {\bf 81}, 115129 (2010).
\bibitem{kiv90} S. A. Kivelson and D. S. Rokhsar, Phys. Rev. B {\bf 41}, 11693 (1990).
\bibitem{lee98} X.-G. Wen and P. A. Lee, Phys. Rev. Lett. {\bf 80}, 2193 (1998).
\bibitem{shar03} S. G. Sharapov, V. P. Gusynin and H. Beck, Phys. Rev. B {\bf 67}, 144509 (2003).
\bibitem{kim02} W. Kim and J. Carbotte, Phys. Rev. B {\bf 66}, 033104 (2002).
\bibitem{vale07} B. Valenzuela and E. Bascones, Phys. Rev. Lett.
{\bf 98}, 227002 (2007).
  \bibitem{notedurs} For example, from Eq. \ref{vheat} the heat current in the electron basis $\hat{U}{\bf v}_{th}\hat{U}$ is $\frac{d}{d{\bf k}}\hat{H}_{\bf k}$, which can be rewritten in the BCS case as ${\bf v}_{th}=\hat{\tau}_{3}{\bf v}_f+\hat{\tau}_{1}{\bf v}_2$ where $\hat{\tau}_\alpha$ is a Pauli matrix.  This is a well known result for $d$-wave superconductors\cite{durs00}.

\end{thebibliography}
\end{document}